\documentclass[draft,twoside]{article}
\newcommand{\mytitle}{The Partition Function of Multicomponent Log-Gases} 
\newcommand{\keywords}{Partition function, Berezin integral, Pfaffian,
Hyperpfaffian, Grand canonical ensemble}

\usepackage{amsmath,amssymb,
  amsthm,amsfonts,amscd,mathrsfs,pifont,upgreek} 
\usepackage{eucal}  
\usepackage{verbatim}
\usepackage{ifpdf}
\usepackage{hyperref}

\ifpdf
        \usepackage[pdftex]{graphicx}
\else
        \usepackage[dvips]{graphicx}
\fi

\newtheorem{thm}{Theorem}[section]
\newtheorem{cor}[thm]{Corollary}
\newtheorem{claim}[thm]{Claim}

\newtheorem{thm*}{Theorem}[]
\newtheorem{cor*}[thm*]{Corollary}
\newtheorem{claim*}[thm*]{Claim}
\newtheorem{lemma*}[thm*]{Lemma}
\newtheorem{prop*}[thm*]{Proposition}
\newtheorem{conj*}[thm*]{Conjecture}

\theoremstyle{definition}

\newtheorem{question*}{Question}[section]

\newtheorem{defn*}{Definition}

\theoremstyle{remark}
\newtheorem*{rem}{Remark}

\newcommand{\qq}[1]{\qquad \mbox{#1} \qquad}

\newcommand{\pdd}[2]{\frac{ \partial {#1} }{ \partial {#2} }}

\newcommand{\BB}[1]{\ensuremath{\mathbb{#1}}}

\newcommand{\R}{\ensuremath{\BB{R}}}

\newcommand{\bs}{\ensuremath{\boldsymbol}}
\newcommand{\mf}{\ensuremath{\mathfrak}}

\newcommand{\prob}{\ensuremath{\mathrm{prob}}}

\newcommand{\qand}{\qquad \mbox{and} \qquad}

\newcommand{\qwhere}{\qquad \mbox{where} \qquad}

\newcommand{\ul}[1]{\underline{#1}}

\DeclareMathOperator{\sgn}{sgn}
\DeclareMathOperator{\vol}{vol}

\DeclareMathOperator{\Pf}{Pf}
\DeclareMathOperator{\PF}{PF}

\numberwithin{equation}{section}

\usepackage{txfonts}
\renewcommand{\Wr}{\mathrm{Wr}}

\begin{document}
\title{\bfseries\sffamily \mytitle}  
\author{\sc Christopher D.~Sinclair\footnote{This research was 
    supported in part by the National Science Foundation (DMS-0801243)} }
\maketitle

\begin{abstract}
  We give an expression for the partition function of a
  one-dimensional log-gas comprised of particles of (possibly)
  different integer charge at inverse temperature $\beta=1$
  (restricted to the line in the presence of a neutralizing field) in
  terms of the Berezin integral of an associated non-homogeneous
  alternating tensor.  This is the analog of the de Bruijn integral
  identities \cite{MR0079647} (for $\beta=1$ and $\beta = 4$)
  ensembles extended to multicomponent ensembles.
\end{abstract}

\vspace{.5cm}
\vspace{.25cm}

\vspace{.5cm}

\section{Introduction}

We imagine a finite number of charged particles interacting
logarithmically on an infinite wire modelled
by the real line.  Different particles may have different charges
(which we will assume are positive integers), but any two particles
with the same charge, that is, of the same {\em species}, are assumed
to be indistinguishable.  A potential is placed on the wire to keep the
particles from escaping to infinity.  This system is placed in contact
with a heat reservoir with inverse temperature $\beta$.   

We will consider two ensembles:
\begin{enumerate}
\item The Canonical Ensemble.  We assume that the number of each
  species of particle is fixed.
\item The Grand Canonical Ensemble.  We assume that the sum of the
  charges, that is the total charge of the system, is fixed but the
  number of each species is variable.\footnote{The standard notion of
    the Grand Canonical Ensemble is that where the number of particles
  is not fixed.  That is, in its traditional sense, the Grand
  Canonical Ensemble is the direct sum over all possible values of the
sum of the charges.  What we refer to as the Grand Canonical Ensemble
might be better referred to as an {\em isocharge} or {\em zero
  current} Grand Canonical Ensemble.} 
\end{enumerate}
Our goal is to provide a closed form of the partition function of
these ensembles, for certain values of $\beta$, in terms of Berezin
integrals.  As is standard, we will find that the partition function
for the Grand Canonical Ensemble is the generating function for the
Canonical ensemble as a function of fugacities of the species of
particles.  After a minor modification, the partition function can
also be seen as the generating function for the correlation functions
of both the Canonical Ensemble and the Grand Canonical Ensemble.  

\subsection{The Setup}

Let $J > 0$ be an integer and suppose $\mathbf q = (q_1, q_2, \ldots, 
q_J)$ is a vector of positive integer (charges) with each of the $q_j$
distinct.  We imagine a system of particles consisting of $M_1$
indistinguishable particles of charge $q_1$, $M_2$ indistinguishable
particles of charge $q_2$ and so on.  We will refer to $\mathbf q$ as
the {\em charge vector} and $\mathbf M = (M_1, M_2, \ldots, M_J)$ as
the {\em population vector} of the system. 

These particles are restricted to lie on an infinite wire, identified
with the real
axis\footnote{With a minor modification, much of what is presented
  here can be shown {\it mutadis mutandis} for multicomponent
  log-gasses confined to the unit circle.  See \cite{MR2450377} for
  the circular case for $\beta=2$, and \cite{MR2450711} for some
  physical application.},
and interact logarithmically, so that the energy contributed to the
system by a pair particles with charges $q$ and $q'$ located at $x$
and $x'$ is given by$-q q' \log|x - x'|$.  (Infinite energy is allowed
in the situation where $x = x'$).  We suppose that the particles of
charge $q_1$ are identified with the location vector
$\mathbf x^1 = (x^1_1, x^1_2, \ldots, x^1_{M_1})$; the location
vectors $\mathbf x^2, \ldots, \mathbf x^J$ are similarly defined.  If
$M_j = 0$ for some $j$ then $\mathbf x^j$ is taken to be the empty
vector.  The particles are placed in a neutralizing field with
potential $U$ so that the total potential energy of the system is
given by
\[
E_{\mathbf M}(\mathbf x^1, \mathbf x^2, \ldots, \mathbf x^J) =
\sum_{j=1}^J q_j \sum_{m=1}^{M_j} U(x^j_m) - \sum_{j=1}^J q_j^2 \sum_{m <
  n} \log| x^j_n - x^j_m | - \sum_{j < k} q_j q_k \sum_{m=1}^{M_j}
\sum_{n=1}^{M_k} \log| x_n^k - x_m^j |.
\]

We assume that the system is in contact with a heat reservoir at
inverse temperature $\beta$, but energy is allowed to flow between the
reservoir and the system of particles.  In this situation the Boltzmann
factor, which gives the relative density of states, is
given by
\begin{align}
\label{eq:4}
& \Omega_{\mathbf M}(\mathbf x^1, \mathbf x^2, \ldots, \mathbf x^J) =
e^{-\beta E_{\mathbf M}(\mathbf x^1, \ldots, \mathbf x^J)}  \\
& \qquad = \bigg\{\prod_{j=1}^J \prod_{m=1}^{M_j} e^{-\beta q_j U(x^j_m)} \bigg\}
\times \bigg\{ \prod_{j=1}^J \prod_{m <
  n} | x^j_n - x^j_m |^{\beta q_j^2} \bigg\} \times \bigg\{
\prod_{j < k} \prod_{m=1}^{M_j} \prod_{n=1}^{M_k} | x_n^k - x_m^j
|^{\beta q_j q_k} \bigg\}.  \nonumber 
\end{align}
The probability (density) of finding the system in a state determined
by the location vectors $\mathbf x^1$, $\mathbf x^2$, $\ldots$, $\mathbf
x^J$ is then given by  
\[
p_{\mathbf M}(\mathbf x^1, \mathbf x^2,
\ldots, \mathbf x^J) = \frac{\Omega_{\mathbf M}(\mathbf
  x^1, \mathbf x^2, \ldots, \mathbf x^J) }{  Z_{\mathbf M} \, M_1! M_2! \cdots M_J!}, 
\] 
where the {\em partition function} of the system is given by 
\begin{equation}
\label{eq:7}
Z_{\mathbf M} = \frac{1}{M_1! M_2! \cdots M_J!}\int_{\R^{M_1}} \cdots
\int_{\R^{M_J}} \Omega_{\mathbf M}(\mathbf
  x^1, \mathbf x^2, \ldots, \mathbf x^J) \, d\mu^{M_1}(\mathbf x^1) \,
  d\mu^{M_2}(\mathbf x^2) \cdots \,
  d\mu^{M_J}(\mathbf x^J),
\end{equation}
and $\mu^M$ is Lebesgue measure on $\R^M$.\footnote{If $M_j = 0$ for some $j$
then we will use the convention that
\[
\int_{\R^{M_j}} \Omega_{\mathbf M}(\mathbf
  x^1, \mathbf x^2, \ldots, \mathbf x^J) \, d\mu^{M_j}(\mathbf x^j) =
  \Omega_{\mathbf M}(\mathbf x^1, \mathbf x^2, \ldots, \mathbf x^J);
\]
alternately, in this situation we may assume that the integral over
$\R^{M_j}$ does not actually appear in our expression for $Z_{\mathbf
  M}$. Likewise we will assume that sums and products over 
empty sets are respectively taken to be 0 and 1.}  The factors of
$M_1! M_2! \cdots M_J!$ appear since a generic state of the system $\mathbf
x^1, \mathbf x^2, \ldots, \mathbf x^J$ has  this many different
representatives.  We will always assume that the external potential
$U$ is such that $Z_{\mathbf M}$ is finite.

For the grand canonical ensemble, We may view $\mathbf M$ as a random
vector and the probability (density) of finding the system with a
prescribed population vector $\mathbf M$ and state
$\mathbf x^1, \mathbf x^2, \ldots, \mathbf x^J$ is given by
$p_{\mathbf M}(\mathbf x^1, \mathbf x^2 \ldots, \mathbf x^J) \cdot
\prob (\mathbf M)$.
Classically, the probability of finding the system in a state with
(allowed) population vector $\mathbf M$ is taken to be
\begin{equation}
\label{eq:12}
\prob(\mathbf M) = z_{1}^{M_1} z_{2}^{M_2} \ldots z_{J}^{M_J}
\frac{Z_{\mathbf M}}{Z_{N}}, 
\end{equation}
where
\[
Z_{N} = \sum_{\mathbf M \atop \mathbf M \cdot \mathbf q = N}
z_{1}^{M_1} z_{2}^{M_2}\ldots z_{J}^{M_1} Z_{\mathbf M} 
\]
and $\mathbf z = (z_{1}, \ldots, z_{J})$ is a vector of positive
real numbers called the {\em fugacity vector}.  

It shall sometimes be convenient to view $\mathbf z$ as a vector of
indeterminants and $Z_{N} = Z_{N}(\mathbf z)$ as a polynomial in these
indeterminants.  Our main result will be to show that, for certain
values of $\beta$ and $\mathbf q$, $Z_{N}(\mathbf z)$ can be expressed
as a Berezin integral with respect to the volume form in $\R^N$ of the
exponential of an (explicitly given) alternating element (i.e.
{\em form}) in the exterior algebra $\Lambda(\R^N)$.  By construction,
$Z_{\mathbf M}$ is the coefficient of $z_{1}^{M_1} \cdots
z_{J}^{M_J}$,
and thus the integral formulation of $Z_{N}(\mathbf z)$ is exactly the
generating function we seek.

\section{Wronskians, Berezin Integrals and Hyperpfaffians}

Here we collect the machinery necessary to state our main results.

Given a non-negative integer $L$, let $\ul L = \{ 1, 2, \ldots L\}$,
and, assuming $K \geq L$ is an integer, let 
\[
\mf t: \ul L \nearrow \ul K
\]
be a strictly increasing function,
\[
0 < \mf t(1) < \mf t(2) < \cdots < \mf t(L) \leq K.
\]  
We will use such functions to keep
track of minors of matrices, elements in exterior algebras and
Wronskians of families of polynomials.  Such indexing functions will
always be written as fraktur minuscules. 

\subsection{Wronskians}

A {\em complete} family of monic polynomials is a sequence of
polynomials $\mathcal P = \left( p_1, p_2, \ldots \right)$ such that
each $p_n$ is monic and $\deg p_n = n-1$.   We define the $L$-tuple 
$\mathcal P_{\mf t} = (p_{\mf t(1)}, \ldots, p_{\mf t(L)})$.  And, 
given $0 \leq \ell < L$ we define the modified $\ell$th
differentiation operator by 
\begin{equation}
\label{eq:1}
D^0 f(x) = f(x) \qand D^{\ell}f(x) = \frac{1}{\ell!}
\frac{d^{\ell}f}{dx^{\ell}}.
\end{equation}
The
{\em Wronskian} of $\mathcal P_{\mf t}$ is then defined to be
\[
\Wr(\mathcal P_{\mf t}; x) = \det \left[ D^{\ell-1} p_{\mf t(k)}(x)
\right]_{k,\ell=1}^L. 
\]
The Wronskian is often defined without the $\ell!$ in the denominator
of~(\ref{eq:1}); this combinatorial factor will prove convenient in
the sequel.  The reader has likely seen Wronskians in elementary
differential equations, where they are used to test for linear
dependence of solutions.   

\subsection{The Berezin Integral}

If $\mathbf e_1, \ldots, \mathbf e_K$ is a
basis for $\R^K$, then $\epsilon_{\mf t} = \mathbf e_{\mf t(1)} \wedge
\cdots \wedge \mathbf e_{\mf t(L)}$ is an element in
$\Lambda^L(\R^K)$, and $\{ \epsilon_{\mf t} \; \big| \; \mf t : \ul L
\nearrow \ul K \}$ is a basis for $\Lambda^L(\R^K)$.  In particular,
we will denote 
\[
\epsilon_{\vol} = \mathbf e_1 \wedge \mathbf e_2 \wedge \cdots \wedge
\mathbf e_K. 
\]

Given $0 < k \leq K$ we define the linear operator $\partial/\partial
\mathbf e_k: \Lambda^L(\R^K) \rightarrow \Lambda^{L-1}(\R^K)$ by 
\[
\frac{\partial}{\partial \mathbf e_k} \epsilon_{\mf t} = \left\{ 
\begin{array}{ll}
(-1)^{\alpha} \, \mathbf e_{\mf t(1)} \wedge \cdots \wedge \mathbf e_{\mf
  t(\alpha - 1)} \wedge \mathbf e_{\mf
  t(\alpha + 1)} \wedge \cdots \wedge \mathbf e_{\mf t(L)}  & \quad
\mbox{if } \; k = \mf t^{-1}(\alpha); \\ & \\
0 & \quad \mbox{otherwise.}
\end{array}
\right.
\]
That is, if $\mathbf e_k$ appears in $\epsilon_{\mf t}$ then $\partial
\epsilon_{\mf t}/\partial \mathbf e_k$ is formed by shuffling $\mathbf
e_k$ to the front of $\epsilon_{\mf t}$ (taking into account the
alternation of signs) and then dropping it.
Given $0 < k_1, \ldots, k_M \leq K$ we then define the Berezin
integral as the linear operator on $\Lambda(\R^K) \rightarrow
\Lambda(\R^K)$ specified by 
\[
\int \epsilon_{\mf t} \, d\mathbf e_{k_1} \, d\mathbf e_{k_2} \, \cdots \, d\mathbf
e_{k_M} = \frac{\partial}{\partial \mathbf e_{k_M}} \cdots 
\frac{\partial}{\partial \mathbf e_{k_2}} \frac{\partial}{\partial
  \mathbf e_{k_1}} \epsilon_{\mf t}. 
\]
Berezin integrals were introduced in \cite{MR0208930} as a Fermionic
analog to the Gaussian integrals which appear in Bosonic field
theory.  

We will mostly be interested in Berezin integrals of the form 
\[
\int \epsilon_{\mf t} \, d\epsilon_{\vol} = \int \epsilon_{\mf t} \,
d\mathbf e_{1} \, \cdots \, d\mathbf e_{K}.
\]
In this case, the Berezin integral is
simply the projection operator $\Lambda(\R^K) \rightarrow
\Lambda^K(\R^K) \cong \R$.  Notice in particular that, if $\sigma \in
S_K$ then  
\[
\int \mathbf e_{\sigma(1)} \wedge \mathbf e_{\sigma(2)} \wedge \cdots
\wedge \mathbf e_{\sigma(K)} \, d\epsilon_{\vol} = \sgn \sigma.
\]

\subsection{Exponentials of Forms and Hyperpfaffians}

Given $\omega \in \Lambda(\R^K)$ we define $\omega^{\wedge 0} = 1$ and for $0 <
m$ 
\[
\omega^{\wedge m} = \underbrace{\omega \wedge \cdots \wedge \omega}_m.
\]
Using this we define
\[
e^{\omega} = \sum_{m=0}^{\infty} \frac{\omega^{\wedge m}}{m!}.
\]
If $\omega = \omega_0 + \omega_1 + \cdots + \omega_K$ with
$\omega_k \in \Lambda^k(\R^K)$ then it is easily verified that
\[
e^{\omega} = e^{\omega_0} \wedge e^{\omega_1} \wedge \cdots \wedge e^{\omega_K}.
\]
Moreover, $e^{\omega_0}$ is a real number equal to its traditional
definition, and if $k > 0$ then the sum defining $e^{\omega_k}$ is a
finite sum. 

In the situation where $k$ divides $K$, that is $K = km$, then we
define the {\em hyperpfaffian} $\PF(\omega_k)$ to be the real number
defined by 
\[
\frac{\omega_k^{\wedge m}}{m!} = \PF(\omega_k) \, \epsilon_{\vol}.
\]
Alternately, 
\[
\PF(\omega_k) = \int e^{\omega_k} \, \epsilon_{\vol}.
\]

The hyperpfaffian is related to the Pfaffian of an antisymmetric $K
\times K$ matrix by associating the matrix to a 2-form in the obvious
manner.  We see therefore that the Berezin integral formed with
respect to $\epsilon_{\vol}$ is a generalization of hyperpfaffians,
which themselves are generalizations of Pfaffians.

\section{Statement of Results}

Suppose $b$ is a positive integer and $\beta = b^2$.  Set $K = bN$
and $L_j = b q_j$ for $j=1,2,\ldots, J$.  For any complete family of
monic polynomials $\mathcal P$ we define $\omega_1,
\omega_2, \ldots, \omega_J \in \Lambda(\R^K)$ as follows.  
\begin{enumerate}
\item If $L_j$ is even, 
\begin{equation}
\label{eq:11}
\omega_j = \sum_{\mf t: \ul L_j \nearrow  \ul K} 
\bigg\{ \int_{\R} e^{-\beta q_j U(x)} \Wr(\mathcal P_{\mf  t}; x) \,
dx 
\bigg\} \epsilon_{\mf t};  
\end{equation}
\item If $L_j$ is odd,
\begin{equation}
\label{eq:3}
\omega_j = \sum_{\mf t,u: \ul L_j \nearrow
  \ul K} \bigg\{\frac12 \int_{\R} \int_{\R} \! e^{-\beta q_j U(x)}e^{-
  \beta q_j U(y)} \Wr(\mathcal P_{\mf   t}; x)
\Wr(\mathcal P_{\mf   u}; y) \sgn(y-x) \, dx dy
\bigg\} \epsilon_{\mf t} \wedge \epsilon_{\mf u}.
\end{equation} 
\end{enumerate}
Notice that $\omega_j$ is in $\Lambda^{L_j}(\R^K)$ when $L_j$ is even
and is in $\Lambda^{2L_j}(\R^K)$ when $L_j$ is odd.  

\begin{thm}
\label{thm:1}
Suppose $\beta = b^2$ and $K = bN$ is even.  Given a charge vector
$\mathbf q$ let  
\[
L_j = b q_j; \qquad j=1,2,\ldots, J,
\]
and, for any complete family of monic polynomials, define the form
$\omega \in \Lambda(\R^{b N})$ by 
\[
\omega(\mathbf z) = \sum_{j=1}^J z_j \omega_{j},
\]
where $\omega_j$ is defined as in (\ref{eq:11}) or (\ref{eq:3}).  
If the $L_j$ are positive integers, at most one of which is odd, then 
\[
Z_{N}(\mathbf z) = \int e^{\omega(\mathbf z)}  \, d\epsilon_{\vol}.
\]
\end{thm}

\begin{rem}
This is an algebraic identity which can be written more generally by
replacing the integral over $\R$ with integrals over other sets (for
instance, the partition functions for multicharge circular ensembles
can be likewise expressed in terms of Berezin integrals).  The only
analytic prerequisite is the finiteness of the $Z_{\mathbf M}$ which
allows for the use of Fubini's Theorem.  
\end{rem}

This theorem covers certain situations which have appeared before.
Certainly the Pfaffian partition functions of the classical
one-species ensembles GOE and GSE  (and their non-Gaussian variants) 
are a corollary.  These cases follow from the `classical' de Bruijn
identities \cite{MR0079647}. (see \cite{MR2129906} and the references contained
therein for their applications to random matrix theory).  The
classical ensembles can be viewed either as ensembles of 
charge 1 particles at respective inverse temperatures $\beta = 1$ and
$\beta = 4$, or to ensembles of charge 1 and charge 2 particles
(respectively) at inverse temperature $\beta = 1$.  Recent work by the
author, B.~Rider and Y.~Xu produced (among other things) a Pfaffian
formulation of the partition function for the grand canonical ensemble
for charge vector $\mathbf q = (1, 2)$ at inverse temperature $\beta =
1$ \cite{rsx}.  When the external field is Gaussian, and by tuning the
fugacity, this latter ensemble provides an unusual interpolation
between the classical ensembles GOE and GSE.  Moreover, the
skew-orthogonal polynomials necessary to solve the ensemble (that is
explicitly derive the matrix kernel in which the correlation functions
can be expressed and analyzed) were explicitly computed in terms of
certain generalized Laguerre polynomials.  Amongst other results, this
allowed us to compute the distribution of the number of each type of
particle for various fugacities.  This analysis follows similar work
for the two charge circular ensemble with charge vector $\mathbf q =
(1,2)$ initiated by P.~Forrester and others (see \cite[\S
7.10]{forrester-book} and the references therein), and indeed the
partition functions for those ensembles can be expressed as a
Pfaffian, and hence in terms of (variants) of the above Berezin
integrals.  

Recent work of the author \cite{Sinclair:2010fk} has lead to a
hyperpfaffian expression for the partition functions of single-species
ensembles of charge 1 particles when $\beta = L^2$ is a perfect
square, or $\beta = L^2 + 1$ is even.  In the former case, these
ensembles can also be interpreted as systems of charge $L$ particles
at $\beta = 1$.  

\subsection{Correlation Functions}

Using a slight modification, the partition function gives a generating
function for the correlation functions.  For single species ensembles,
the correlation functions are simply renormalized marginal densities.  For
multicomponent ensembles, however, the situation is more complicated
(though the marginal probabilities are an important ingredient). 

For fixed population vector $\mathbf M$ and vector $\mathbf m = (m_1,
m_2, \ldots, m_J)$ with $0 \leq m_j \leq M$, we define
\[
\bs \xiup^j = (\xi^j_1, \ldots, \xi^j_{m_j}) \qq{and} \mathbf y^j =
(y^j_1, \ldots, y^j_{M_j - m_j}),
\]
and set
\[
\bs \xiup^j \vee \mathbf y^j = (\xi^j_1, \ldots, \xi^j_{m_j}, y^j_1,
\ldots, y^j_{M_j - m_j}).
\]

The $\mathbf m$th marginal probability density of $p_{\mathbf M}$ is then
given by
\[
p_{\mathbf M, \mathbf m}(\bs \xiup^1, \bs \xiup^2, \ldots, \bs
\xiup^J) = \int_{\R^{M_1 - m_1}} \cdots \int_{\R^{M_J - m_J}}
p_{\mathbf M}(\bs \xiup^1 \vee \mathbf y^1, \ldots, \bs \xiup^J \vee
\mathbf y^J ) \, d\mu^{M_1 - m_1}(\mathbf y^1) \cdots d\mu^{M_J -
  m_j}(\mathbf y^J), 
\]
and by symmetry, the probability (density) that our system is in a
state $(\mathbf x^1, \ldots, \mathbf x^J)$ which occupies the substate
$(\bs \xiup_1, \ldots, \bs \xiup_J)$ (that is, viewed as sets, $\bs
\xiup^j \subseteq \mathbf x^j$ for each $j$) is given by
\begin{align}
&R_{\mathbf M, \mathbf m}(\bs \xiup^1, \bs \xiup^2, \ldots, \bs
\xiup^J) = \frac{M_1!}{(M_1 - m_1)!} \cdots \frac{M_J!}{(M_J - m_J)!}
p_{\mathbf M, \mathbf m}(\bs \xiup^1, \bs \xiup^2, \ldots, \bs
\xiup^J)  \nonumber \\
& \qquad = \frac{1}{Z_{\mathbf M} \, (M_1 - m_1)! \cdots (M_J -
  m_J)!} \nonumber \\ 
& \hspace{1.5cm} \times \int_{\R^{M_1 - m_1}} \cdots \int_{\R^{M_J - m_J}}
\Omega_{\mathbf M}(\bs \xiup^1 \vee \mathbf y^1, \ldots, \bs \xiup^J \vee
\mathbf y^J ) \, d\mu^{M_1 - m_1}(\mathbf y^1) \cdots d\mu^{M_J - 
  m_j}(\mathbf y^J). \label{eq:13} 
\end{align}
This is the $\mathbf m$th {\em correlation function} for the canonical
ensemble with population vector $\mathbf M$.   

To get the $\mathbf m$th correlation function for the grand
canonical ensemble we need to sum over the related correlation
function for the canonical ensemble over all allowable population vectors
$\mathbf M$ with $m_j \leq M_j$ for each $j$ (a situation we will
abbreviate by $\mathbf m \leq \mathbf M$), taking into account the
probability of being in a state with prescribed
population vector.  That is, the probability (density) of the (grand canonical)
system is in a state $(\mathbf x^1, \ldots, \mathbf x^J)$ which
occupies the substate $(\bs \xiup_1, \ldots, \bs \xiup_J)$ is given by
\[
\prob\big( (\bs \xiup_1, \ldots, \bs \xiup_J) \subseteq (\mathbf x^1,
\ldots, \mathbf x^J) \big) = \sum_{\mathbf M.\mathbf q = N \atop
  \mathbf M \geq \mathbf m} \prob(\mathbf M) \cdot R_{\mathbf M, \mathbf
  m}(\bs \xiup^1, \bs \xiup^2, \ldots, \bs \xiup^J).
\]
Denoting this density by $R_{N, \mathbf m}$, (\ref{eq:12}) and (\ref{eq:13}) yield
\begin{align*}
& R_{N, \mathbf m}(\bs \xiup_1, \ldots, \bs \xiup_J) =
\frac{1}{Z_N(\mathbf z)} \sum_{\mathbf M.\mathbf q = N \atop \mathbf M
  \geq \mathbf m} \frac{z_1^{M_1} \cdots z_J^{M_J}}{(M_1 - m_1)!
  \cdots (M_J - m_j)!} \\
& \hspace{1.5cm} \times \int_{\R^{M_1 - m_1}} \cdots \int_{\R^{M_J - m_J}}
\Omega_{\mathbf M}(\bs \xiup^1 \vee \mathbf y^1, \ldots, \bs \xiup^J \vee
\mathbf y^J ) \, d\mu^{M_1 - m_1}(\mathbf y^1) \cdots d\mu^{M_J - 
  m_j}(\mathbf y^J).
\end{align*}
Notice that, by ignoring the prefactor $Z_N(\mathbf z)$, and up to an
easily recoverable constant, the coefficient of $z_1^{M_1} \cdots
z_J^{M_J}$ in $R_{N,\mathbf m}$ is the $\mathbf m$th correlation
function for the corresponding canonical ensemble.  

We can in turn give a generating function for the correlation
functions for the grand canonical ensemble as follows:  Let
$\mathbf c^j = (c^j_1, c^j_2, \ldots, c^j_N)$ and $\bs \zetaup^j =
(\zeta_1^j, \zeta_2^j, \ldots, \zeta_N^j)$ and define the measures
$\nu_j$ and $\eta_j$ by 
\[
\frac{d \nu_j}{d\mu_j}(x) = e^{-\beta q_j U(x)} \qq{and} 
\eta_j(x) = e^{-\beta q_j U(x)} \sum_{n=1}^N c_n^j \delta(x - \zeta_n^j),
\]
where $\delta(x)$ is the probability measure with unit mass at $x=0$.  

It is convenient at this point to index the forms $\omega_j$ from
(\ref{eq:11}) and (\ref{eq:3}) by $\nu_j$ so that, for instance when
$L_j$ is even, 
\[
\omega^{\nu_j}_j = \sum_{\mf t: \ul L_j \nearrow \ul K} \bigg\{ \int_{\R}
\Wr(\mathscr P_{\mf t}) \, d\nu_j \bigg\} \epsilon_{\mf t}.
\]
Quantities which are dependent on $\bs \nuup = (\nu_1, \ldots,
\nu_J)$ will be denoted by, for instance, $Z_{\mathbf M}^{\bs \nuup},
Z_N^{\bs \nuup}$ and $\omega^{\bs \nuup}(\mathbf z)$.  Theorem
\ref{thm:1} is purely algebraic, and thus, we have, for instance that
\[
Z_N^{\bs \nuup}(\mathbf z) = \int \exp\{\omega^{\nuup}(\mathbf z)\} \, d\epsilon_{\vol}.
\]
We can generalize these quantities by replacing $\bs \nuup$ with other
vectors of measures.  The following theorem gives particular relevance
to $Z_N^{\bs \nuup + \bs \etaup}(\mathbf z, \mathbf c^1, \ldots,
\mathbf c^J)$, where the
notation indicates the additional dependence on the $\mathbf c^j$.  
\begin{claim}
\label{claim:1}
The $\mathbf m$th correlation function of the grand canonical ensemble is the
coefficient of 
\[
\prod_{j=1}^J \prod_{\ell=1}^{m_j} c_{\ell}^j \qq{in} \frac{Z_N^{\bs
    \nuup + \bs \etaup}(\mathbf z, \mathbf c^1, \ldots, \mathbf
  c^J)}{Z_N(\mathbf z)}.
\]
That is, if $\bs \xiup^j = (\zeta_1^j, \ldots, \zeta_{m_j}^j)$, and we
define 
\[
\pdd{^{m_j}}{\mathbf c^j} = \pdd{}{c_{1}^{j}} \cdots \pdd{}{c_{m_j}^{j}}
\]
then,
\[
R_{N, \mathbf m}(\bs \xiup^1, \ldots, \bs \xiup^J) = \frac{1}{Z_N(\mathbf
z)} \left[\pdd{^{m_1}}{\mathbf c^1} \cdots \pdd{^{m_J}}{\mathbf c^J} Z_N^{\bs
    \nuup + \bs \etaup}(\mathbf z, \mathbf c^1, \ldots, \mathbf
  c^J) \right|_{\mathbf c^1 = \cdots = \mathbf c^J =
  \bs 0}.
\]
Moreover, the $\mathbf m$th correlation function of the canonical
ensemble with population vector $\mathbf M$ is given by 
\[
R_{\mathbf M, \mathbf m} (\bs \xiup^1, \ldots, \bs \xiup^J) =
\frac{1}{Z_{\mathbf M} M_1! \cdots M_J!} \pdd{^{M_1}}{z_1^{M_1}} \cdots
\pdd{^{M_J}}{z_J^{M_J}} \left[\pdd{^{m_1}}{\mathbf c^1} \cdots
  \pdd{^{m_J}}{\mathbf c^J}Z_N^{\bs 
    \nuup + \bs \etaup}(\mathbf z, \mathbf c^1, \ldots, \mathbf
  c^J) \right|_{\mathbf z = \mathbf c^1 = \cdots = \mathbf c^J =
  \bs 0}.
\]
\end{claim}

The proof of this claim is standard (it is the multicomponent
version of the `functional differentiation' method), and follows {\em mutatis
  mutandis} that for Ginibre's real ensemble
\cite[Prop.~6]{borodin-2008}.

To write the correlation functions explicitly in terms of a Berezin
integral (taking all of the $L_j$ to be even for convenience), we note that
\[
\omega_j^{\nu_j + \eta_j} = \sum_{\mf t: \ul L_j \nearrow \ul K} \bigg\{ \int_{\R}
\Wr(\mathscr P_{\mf t}) \, d(\nu_j + \eta_j) \bigg\} \epsilon_{\mf t}
= \omega_j^{\nu_j} + \omega_j^{\eta_j},
\]
and
\[
\omega^{\bs \nuup + \bs \etaup} = \sum_{j=1}^J \omega_j^{\nu_j +
  \eta_j} = \omega^{\bs \nuup} +
\omega^{\bs \etaup},
\]
Hence, $\exp\{\omega^{\bs \nuup + \bs \etaup}\} = \exp\{\omega^{\bs
  \nuup}\} \wedge \exp\{\omega^{\bs \etaup}\}.$

This is useful, since the first term in the right hand side is
independent of the $\mathbf c^j$.  The following maneuvers are elementary
\begin{align*}
\pdd{^{m_1}}{\mathbf c^1}  \exp\{\omega^{\bs
  \nuup + \bs \etaup}\} &=  \exp\{\omega^{\bs
  \nuup}\} \wedge \pdd{^{m_1}}{\mathbf c^1} \exp\{\omega^{\bs \etaup}\} \\
&= \exp\{\omega^{\bs \nuup}\} \wedge \exp\{\omega^{\bs \etaup}\}
\wedge 
\bigg\{ \bigwedge_{\ell=1}^{m_1} e^{-\beta q_1
  U(\zeta^1_{\ell})} \sum_{\mf t: \ul L_1 \nearrow \ul K}\Wr(\mathcal
P_{\mf t}; \zeta^1_{\ell}) \epsilon_{\mf t} \bigg\}; 
\end{align*}
note that since all forms are even, we do not have to specify their
order.  It follows that
\[
\pdd{^{m_1}}{\mathbf c^1}  \exp\{\omega^{\bs
  \nuup}\}\wedge \exp\{\omega^{\bs \etaup}\} \bigg|_{\mathbf c^1 = 0}
=  \exp\{ \omega^{\nu} \} \wedge \bigg\{ \bigwedge_{\ell=1}^{m_1} e^{-\beta q_1
  U(\zeta^1_{\ell})} \sum_{\mf t: \ul L_1 \nearrow \ul K} \Wr(\mathcal
P_{\mf t}; \zeta^1_{\ell}) \epsilon_{\mf t}\bigg\}, 
\]
and that
\[
\pdd{^{m_1}}{\mathbf c^1} \cdots \pdd{^{m_J}}{\mathbf c^J} \exp\{\omega^{\bs \nuup}\}\wedge \exp\{\omega^{\bs \etaup}\} \bigg|_{\mathbf c^1 = \cdots = \mathbf c^J =
  \bs 0} = \exp\{ \omega^{\nu} \} \wedge \bigg\{ \bigwedge_{j=1}^J 
\bigwedge_{\ell=1}^{m_j} e^{-\beta q_j U(\zeta^j_{\ell})} \sum_{\mf t:
  \ul L_j \nearrow \ul K}\Wr(\mathcal
P_{\mf t}; \zeta^j_{\ell}) \epsilon_{\mf t} \bigg\}.
\]
We therefore have the following corollary to Claim~\ref{claim:1}.
\begin{cor}
If, for each $1 \leq j \leq J$, $L_j$ is even, and $\bs \xiup^j =
(\zeta_1^j, \ldots, \zeta_{m_j}^j) \in \R^{m_j}$, then
\[
R_{N, \mathbf m}(\bs \xiup^1, \ldots, \bs \xiup^J) =
\frac{1}{Z_N(\mathbf z)}
\int \exp\{ \omega^{\nu} \} \wedge \bigg\{ \bigwedge_{j=1}^J 
\bigwedge_{\ell=1}^{m_j} e^{-\beta q_j U(\zeta^j_{\ell})} \sum_{\mf t:
  \ul L_j \nearrow \ul K}\Wr(\mathcal
P_{\mf t}; \zeta^j_{\ell}) \epsilon_{\mf t} \bigg\} \, d\epsilon_{\vol}.
\]
\end{cor}
Note that we do not have to justify the exchange of the derivatives
and the `integral' in Claim~\ref{claim:1}, since the Berezin integral
is not an integral in the traditional sense.  That is,
Claim~\ref{claim:1} is an algebraic, not an analytic, identity.
Notice also, that the quantity in braces is an $(\mathbf m \cdot \mathbf
L)$-form, and therefore only the projection of $\exp{\omega^{\nu}}$
onto the space of $(K - \mathbf m \cdot \mathbf L)$-forms will make a
contribution to the $\mathbf m$th correlation function.  Finally, we
note that a similar formula for the partial correlation function
$R_{\mathbf M, \mathbf m}$ is available via functional differentiation with respect
to the $z$ variables.  

\section{The Proof of Theorem~\ref{thm:1}} 

\subsection{The Confluent Vandermonde Determinant}

Suppose $0 < L < K$, $x \in \R$ and $\mathcal P = (p_1, p_2, \ldots)$ is any
complete family of monic polynomials. We define the $K \times L$ matrix  
\[
\mathbf V^{L}(x) = \left[ D^{\ell} p_n(x) \right]_{n,\ell=1}^{K, L},
\]
and given an admissible population vector $\mathbf M$ and $\mathbf
x^1, \ldots, \mathbf x^J$ with $\mathbf x^j \in \R^{M_j}$, we define
the $K \times K$ confluent Vandermonde matrix by  
\[
\mathbf V^{\mathbf M}(\mathbf x^1, \ldots, \mathbf x^J) = 
\bigg[
\underbrace{\mathbf V^{L_1}(x_1^1) \quad \cdots \quad \mathbf 
  V^{L_1}(x_{M_1}^1)}_{M_1} \qquad \cdots \qquad \underbrace{\mathbf
  V^{L_J}(x_1^J) \quad \cdots \quad \mathbf  V^{L_J}(x_{M_J}^J)}_{M_J}.
\bigg]
\]
(Recall that $L_j = \sqrt{\beta} q_j$).  
In this case, the confluent Vandermonde determinant identity
\cite{Meray} has that 
\begin{align}
\det \mathbf V^{\mathbf M}(\mathbf x^1, \ldots, \mathbf
x^J)
 &= \bigg\{ \prod_{j=1}^J \prod_{m <
  n} ( x^j_n - x^j_m )^{L_j^2} \bigg\} \times \bigg\{
\prod_{j < k} \prod_{m=1}^{M_j} \prod_{n=1}^{M_k} ( x_n^k - x_m^j
)^{L_j L_k} \bigg\} \label{eq:9} \\
&= \bigg\{ \prod_{j=1}^J \prod_{m <
  n} ( x^j_n - x^j_m )^{b^2 q_j^2} \bigg\} \times \bigg\{
\prod_{j < k} \prod_{m=1}^{M_j} \prod_{n=1}^{M_k} ( x_n^k - x_m^j
)^{b^2 q_j q_k} \bigg\}. \nonumber 
\end{align}
When all of the $L_j$ are even, it follows from (\ref{eq:4}) that
\begin{equation}
\label{eq:5}
\Omega_{\mathbf M}(\mathbf x^1, \mathbf x^2, \ldots,
\mathbf x^J) = \bigg\{\prod_{j=1}^J \prod_{m=1}^{M_j} e^{-\beta
q_jU(x^j_m)} 
\bigg\} \det \mathbf V^{\mathbf M}(\mathbf x^1, \ldots, \mathbf
x^J)
\end{equation}
We will deal with the situation where one of the $L_j$ is odd in
Section~\ref{sec:when-one-charges}.

\subsection{The Laplace Expansion of the Determinant}

Each $\mf t: \ul L \nearrow \ul K$ specifies a unique
$\mf t' : \ul{K - L} \nearrow \ul N$ whose range is disjoint from
$\mf t$.
Given a $K \times K$ matrix $\mathbf V = [v_{m,n}]$ and $\mf t, \mf
u: \ul L \nearrow \ul K$ then we may create a $L \times L$ minor of
$\mathbf V$ by selecting the rows and columns from the ranges of $\mf
t$ and $\mf u$.  That is, we write 
\[
\mathbf V_{\mf t, \mf u} = \left[ v_{\mf t(k), \mf u(\ell)} \right]_{k,\ell=1}^L.
\]
Notice that the complementary minor to $\mathbf V_{\mf t, \mf u}$ is
given by $\mathbf V_{\mf t', \mf u'}$.  

We define $\sgn \mf t$ by
\[
 \sgn \mf t = \int \epsilon_{\mf t} \wedge \epsilon_{\mf t'} \, d\epsilon_{\vol}.
\]
More generally let
\[
\mf t_m^j : \ul L_j \rightarrow \ul K \qwhere j=1,2,\ldots J \qand 
  m=1,2,\ldots,M_j,
\]
and set 
\[
\vec{\mf t} = (\underbrace{\mf t_1^1, \ldots, \mf t_{M_1}^1}_{M_1},
\cdots, 
\underbrace{\mf t_1^J, \ldots, \mf t_{M_J}^J}_{M_J} )
\]
We will use $\vec{\mf t}$ to select minors of $\mathbf V^{\mathbf
  M}(\mathbf x^1, \ldots, \mathbf x^J)$ each of which depends 
only on a single location variable.  We denote the set of all such $\vec{
\mf t}$ by $\mf I_{\mathbf m}$ 

We define $\sgn \vec{\mf t}$ by 
\begin{equation}
\label{eq:2}
\sgn \vec{\mf t} = \int \underbrace{\epsilon_{\mf t_1^1} \wedge \cdots
  \wedge \epsilon_{\mf t_{M_1}^1}}_{M_1} \wedge \cdots \wedge 
\underbrace{\epsilon_{\mf t_1^J} \wedge \cdots \wedge \epsilon_{\mf t_{M_J}^J}}_{M_J}
d \epsilon_{\vol}.
\end{equation}
Clearly, $\sgn \vec{\mf t} = 0$ unless the ranges of the various $\mf
t_m^j$ are mutually disjoint, and otherwise $\sgn \mf t$ is the
signature of the permutation defined by concatenating the ranges of
the various $\mf t_m^j$ in the appropriate order.  

We will reserve the symbol $\vec{\mf
  i}$ for the vector whose coordinate functions are given by 
\[
\mf i_m^j(\ell) = \ell + (m-1) L_j + M_1  L_1 + \cdots + M_{j-1}
L_{j-1} 
\]
That is, for instance, if $\mathbf L = (2,3)$ and $\mathbf M = (2,2)$ 
then the ranges of $\mf i_1^1, \mf i_2^1, \mf i_1^2$ and $\mf i_2^2$
are given respectively by $\{1,2\}, \{3,4\}, \{5,6,7\}$ and
$\{8,9,10\}$.  Clearly $\sgn \vec{\mf i} = 1$.  

This notation is convenient to
represent the Laplace expansion of the determinant (which we will
write in the form most useful for our ultimate goal).
\begin{equation}
\label{eq:6}
\det \mathbf V = \sum_{\vec{\mf t} \in \mf I_{\mathbf M}} \sgn \vec{\mf t} \; \prod_{j=1}^J
\prod_{m=1}^{M_j} \det \mathbf V_{\mf t^j_m, \mf i^j_m}.
\end{equation}

Applying (\ref{eq:6}) to $\mathbf V^{\mathbf M}(\mathbf
x^1, \ldots, \mathbf x^J)$ we find
\[
\det \mathbf V^{\mathbf M}(\mathbf
x^1, \ldots, \mathbf x^J) = \sum_{\vec{\mf t}  \in \mf I_{\mathbf M}}
\sgn \vec{\mf t} \;  
\prod_{j=1}^J \prod_{m=1}^{M_j}
\det \mathbf V^{\mathbf M}_{\mf t_m^j, \mf i_m^j}(x_m^j),
\]
where the notation reflects the fact that $\mathbf V^{\mathbf M}_{\mf
  t_{j}, \mf i_j}(x_m^j)$ in independent of all location variables
except $x_m^j$.  From the definition of $\mathbf V^{\mathbf M}$ we see that
$\det \mathbf V^{\mathbf M}_{\mf t_m^j, \mf i_m^j}(x_m^j) =
\Wr(\mathcal P_{\mf t_m^j}; x_m^j)$, and therefore 
\begin{equation}
\label{eq:8}
\det \mathbf V^{\mathbf M}(\mathbf
x^1, \ldots, \mathbf x^J) = \sum_{\vec{\mf t}  \in \mf I_{\mathbf M}}
\sgn \vec{\mf t} \;  
\prod_{j=1}^J \prod_{m=1}^{M_j}
\Wr(\mathcal P_{\mf t_m^j}; x_m^j).
\end{equation}

\subsection{Fubini's Theorem}

From (\ref{eq:7}), (\ref{eq:5}) and (\ref{eq:8}) we have that
\begin{align*}
& Z_{\mathbf M} = \frac{1}{M_1! M_2! \cdots M_J!} \sum_{\vec{\mf t}  \in \mf I_{\mathbf M}}
\sgn \vec{\mf t} \int_{\R^{M_1}} \int_{\R^{M_2}} \cdots
\int_{\R^{M_J}} \prod_{j=1}^J \prod_{m=1}^{M_j}  e^{-\beta
q_j  U(x^j_m)} \Wr(\mathcal P_{\mf t_m^j}; x_m^j) \\
& \hspace{8cm} \times  \, d\mu^{M_1}(\mathbf x^1) \,
  d\mu^{M_2}(\mathbf x^2) \cdots \,
  d\mu^{M_J}(\mathbf x^J).
\end{align*}
Fubini's Theorem implies then that 
\begin{equation}
\label{eq:10}
Z_{\mathbf M} = \frac{1}{M_1! M_2! \cdots M_J!} \sum_{\vec{\mf t}  \in \mf I_{\mathbf M}}
\sgn \vec{\mf t} \; \prod_{j=1}^J \prod_{m=1}^{M_j} \int_{\R} e^{-\beta
q_j  U(x)} \Wr(\mathcal P_{\mf t_m^j}; x) \, dx.
\end{equation}
Thus,
\begin{align*}
Z_N(\mathbf z) &= \sum_{\mathbf M \atop \mathbf M \cdot \mathbf q = N}
\frac{z_1^{M_1} z_2^{M_2} \cdots z_J^{M_J}}{M_1! M_2! \cdots M_J!} \sum_{\vec{\mf t}  \in \mf I_{\mathbf M}}
\sgn \vec{\mf t} \; \prod_{j=1}^J \prod_{m=1}^{M_j} \int_{\R} e^{-\beta
q_j  U(x)} \Wr(\mathcal P_{\mf t_m^j}; x) \, dx \\
 &= \sum_{\mathbf M \atop \mathbf M \cdot \mathbf q = N}
 \sum_{\vec{\mf t}  \in \mf I_{\mathbf M}} 
\sgn \vec{\mf t} \; \prod_{j=1}^J \frac{1}{M_j!} \prod_{m=1}^{M_j} z_j
\int_{\R} e^{-\beta q_j U(x)} \Wr(\mathcal P_{\mf t_m^j}; x) \, dx.
\end{align*}

\subsection{Enter the Berezin Integral}

Using the definition of $\sgn \vec{\mf t}$ (\ref{eq:2}) we find
\begin{align*}
& Z_N(\mathbf z) = \sum_{\mathbf M \atop \mathbf M \cdot \mathbf q = N}
 \sum_{\vec{\mf t}  \in \mf I_{\mathbf M}} \bigg\{ \int
 \underbrace{\epsilon_{\mf t_1^1} \wedge \cdots 
  \wedge \epsilon_{\mf t_{M_1}^1}}_{M_1} \wedge \cdots \wedge 
\underbrace{\epsilon_{\mf t_1^J} \wedge \cdots \wedge \epsilon_{\mf t_{M_J}^J}}_{M_J}
d \epsilon_{\vol} \bigg\} \\ & \hspace{6cm} 
 \times  \prod_{j=1}^J \frac{1}{M_j!} \prod_{m=1}^{M_j} z_j
\int_{\R} e^{-\beta q_j U(x)} \Wr(\mathcal P_{\mf t_m^j}; x) \, dx. 
\end{align*}
Exploiting the linearity of the Berezin integral, 
\[
Z_N(\mathbf z) = \int \bigg[ \sum_{\mathbf M \atop \mathbf M \cdot \mathbf q = N}
 \sum_{\vec{\mf t}  \in \mf I_{\mathbf M}} \bigwedge_{j=1}^J
 \frac{1}{M_j!} \bigwedge_{m=1}^{M_j} z_j 
\bigg\{ \int_{\R} e^{-\beta q_j U(x)} \Wr(\mathcal P_{\mf t_m^j}; x) \, dx
\bigg\} \epsilon_{\mf t_m^j} \bigg] d\epsilon_{\vol},
\]
where the wedge products are taken in the standard order.  

Next we may expand the sum over $\vec{\mf t} \in \mf I_{\mathbf M} $ as
\[
\sum_{\vec{\mf t} \in \mf I_{\mathbf M}} \big( \cdots \big) = \sum_{\mf
  t_1^1, \ldots, \mf t_{M_1}^1: \ul L_1 \nearrow \ul K} \quad \sum_{\mf
  t_1^2, \ldots, \mf t_{M_2}^2: \ul L_2 \nearrow \ul K}  \cdots \sum_{\mf
  t_1^J, \ldots, \mf t_{M_1}^J: \ul L_J \nearrow \ul K}  \big( \cdots \big),
\]
so that 
\[
Z_N(\mathbf z) = \int \bigg[ \sum_{\mathbf M \atop \mathbf M \cdot \mathbf q = N}
\bigwedge_{j=1}^J
 \frac{1}{M_j!} \sum_{\mf
  t_1^j, \ldots, \mf t_{M_j}^j: \ul L_j \nearrow \ul K}
\bigwedge_{m=1}^{M_j} z_j \bigg\{ \int_{\R} e^{-\beta q_j U(x)}
\Wr(\mathcal P_{\mf t_m^j}; x) \, dx 
\bigg\} \epsilon_{\mf t_m^j} \bigg] d\epsilon_{\vol}.
\]
We observe that 
\begin{align*}
&\sum_{\mf
  t_1^j, \ldots, \mf t_{M_j}^j: \ul L_j \nearrow \ul K}
\bigwedge_{m=1}^{M_j} z_j \bigg\{ \int_{\R} e^{-\beta q_j U(x)}
\Wr(\mathcal P_{\mf t_m^j}; x) \, dx 
\bigg\} \epsilon_{\mf t_m^j}  \\
& \hspace{4cm} = \bigg(z_j \sum_{\mf t: \ul L_j \nearrow
  \ul K} \bigg\{ \int_{\R} e^{-\beta q_j U(x)}
\Wr(\mathcal P_{\mf t}; x) \, dx 
\bigg\} \epsilon_{\mf t} \bigg)^{\wedge M_j} \\
& \hspace{4cm} = (z_j \omega_{j})^{\wedge M_j},
\end{align*}
and hence 
\[
Z_N(\mathbf z) = \int \bigg[ \sum_{\mathbf M \atop \mathbf M \cdot \mathbf q = N}
\bigwedge_{j=1}^J
 \frac{(z_j \omega_{j})^{\wedge M_j} }{M_j!} \bigg] d\epsilon_{\vol}.
\]

Now, we can remove the restriction $\mathbf M \cdot \mathbf q = N$
from the sum in this expression, since the Berezin integral will be
zero for any $\mathbf M$ not satisfying this condition.  (If $\mathbf
M$ does not satisfy this condition the form in the integrand will not
be in $\Lambda^K(\R^K)$ and hence its projection onto $\Lambda^K(\R^K)
\cong \R$ will be 0).  Thus,
\begin{align*}
Z_N(\mathbf z) &= \int \bigg[ \sum_{\mathbf M}
\bigwedge_{j=1}^J
 \frac{(z_j \omega_{j})^{\wedge M_j} }{M_j!} \bigg]
 d\epsilon_{\vol} \\
&= \int \bigg[ \sum_{M_1 = 0}^{\infty} \sum_{M_2 = 0}^{\infty} \cdots
\sum_{M_J = 0}^{\infty} 
\bigwedge_{j=1}^J
 \frac{(z_j \omega_{j})^{\wedge M_j} }{M_j!} \bigg]
 d\epsilon_{\vol} \\
&= \int \bigg[
\bigwedge_{j=1}^J \sum_{M=1}^{\infty} 
 \frac{(z_j \omega_{j})^{\wedge M} }{M!} \bigg]
 d\epsilon_{\vol} \\
&= \int 
e^{z_1 \omega_{1}} \wedge e^{z_2 \omega_{2}} \wedge \cdots \wedge
e^{z_J \omega_{j}}  d\epsilon_{\vol} \\
&= \int e^{\omega(\mathbf z)} d\epsilon_{\vol},
\end{align*}
as desired.  

\subsection{When one of the $L_j$ is odd}
\label{sec:when-one-charges}

In the case where exactly one of the $L_j$ is odd, we will reorder the
$q_j$ so that $L_1$ is odd and $L_2, \ldots, L_J$ are even.  In this
situation, (\ref{eq:4}) and (\ref{eq:9}) imply that 
\[
\Omega_{\mathbf M}(\mathbf x^1, \mathbf x^2, \ldots, \mathbf x^J) =
\bigg\{
\prod_{j=1}^J \prod_{m=1}^{M_j} e^{-\beta q_j U(x_m^j)} 
\bigg\} \bigg\{ \prod_{1 \leq m < n \leq M_1} \sgn(x_n^1 - x_m^1)
\bigg\} \det \mathbf V^{\mathbf M}(\mathbf x^1, \mathbf x^2, \ldots,
\mathbf x^J),
\]
where the additional factors of the form $\sgn(x^1_n - x^1_m)$ exist
in order to make the expression non-negative for all choices of
$\mathbf x^1, \mathbf x^2, \ldots, \mathbf x^J$.  Defining the $M_1
\times M_1$ antisymmetric matrix 
\[
\mathbf T(\mathbf x^1) = \left[ \sgn(x^1_n - x^1_m) \right]_{m,n = 1}^{M_1},
\]
When $K$ is even, so is $M_1$, and in this situation 
\[
\Pf \mathbf T(\mathbf x^1) = \prod_{1 \leq m < n \leq M_1} \sgn(x_n^1
- x_m^1).
\]
Thus,
\[
\Omega_{\mathbf M}(\mathbf x^1, \mathbf x^2, \ldots, \mathbf x^J) =
\bigg\{
\prod_{j=1}^J \prod_{m=1}^{M_j} e^{-\beta q_j U(x_m^j)} 
\bigg\} \Pf \mathbf T(\mathbf x^1) \; \det \mathbf V^{\mathbf M}(\mathbf x^1, \mathbf x^2, \ldots,
\mathbf x^J),
\]
Following the analysis of the case where all $L_j$ even we find the
analog of (\ref{eq:10}) in the current situation is
\begin{align*}
&Z_{\mathbf M} = \frac{1}{M_1! M_2! \cdots M_J!} \sum_{\vec{\mf t}  \in \mf I_{\mathbf M}}
\sgn \vec{\mf t} \; \bigg\{  \prod_{j=2}^J \prod_{m=1}^{M_j} \int_{\R}
e^{-\beta q_j  U(x)} \Wr(\mathcal P_{\mf t_m^j}; x) \, dx \bigg\} \\
& \hspace{4cm} \times \int_{\R^{M_1}} \Pf \mathbf T(\mathbf x)
\bigg\{ \prod_{m=1}^{M_1}  e^{-\beta q_1 U(x_m)} \Wr(\mathcal P_{\mf
  t_m^1}; x_m)  \bigg\}
\, d\mu^{M_1}(\mathbf x)
\end{align*}
And,
\begin{align*}
&Z_{N}(\mathbf z) = \sum_{\mathbf M \atop \mathbf M \cdot \mathbf q =
  N} \frac{z_1^{M_1} z_2^{M_2} \cdots z_J^{M_J}}{M_1! M_2! \cdots
  M_J!} \sum_{\vec{\mf t}  \in \mf I_{\mathbf M}} 
\sgn \vec{\mf t} \; \bigg\{  \prod_{j=2}^J \prod_{m=1}^{M_j} \int_{\R}
e^{-\beta q_j  U(x)} \Wr(\mathcal P_{\mf t_m^j}; x) \, dx \bigg\} \\
& \hspace{4cm} \times \int_{\R^{M_1}} \Pf \mathbf T(\mathbf x)
\bigg\{ \prod_{m=1}^{M_1}  e^{-\beta q_1 U(x_m)} \Wr(\mathcal P_{\mf
  t_m^1}; x_m)  \bigg\}
\, d\mu^{M_1}(\mathbf x).
\end{align*}
Using the same maneuvers as before, we can write 
\begin{align*}
&Z_{\mathbf M} = \int \bigg[
 \sum_{\mathbf M \atop \mathbf M \cdot \mathbf q = N}
\bigwedge_{j=2}^J
 \frac{(z_j \omega_{j})^{\wedge M_j} }{M_j!} 
 \wedge \frac{z_1^{M_1}}{M_1!} \sum_{\mf
  t_1^1, \ldots, \mf t_{M_1}^1: \ul L_1 \nearrow \ul K} \\ &
\hspace{2cm} \bigg(
\int_{\R^{M_1}} \Pf \mathbf T(\mathbf x) \bigg\{ \prod_{m=1}^{M_1}
e^{-\beta q_1 U(x_m)} \Wr(\mathcal P_{\mf   t_m^1}; x_m)  \bigg\}
\, d\mu^{M_1}(\mathbf x) \bigg) \epsilon_{\mf t_1^1} \wedge
\epsilon_{\mf t_2^1} \wedge \cdots \wedge \epsilon_{\mf t_{M_1}^1}
\bigg] d\epsilon_{\vol}.  
\end{align*}

It is shown in \cite[Section 4.2]{Sinclair:2010fk} that 
\begin{align*}
& \frac{z_1^{M_1}}{M_1!} \sum_{\mf
  t_1^1, \ldots, \mf t_{M_1}^1: \ul L_1 \nearrow \ul K} 
\bigg(
\int_{\R^{M_1}} \Pf \mathbf T(\mathbf x) \bigg\{ \prod_{m=1}^{M_1}
e^{-\beta q_1 U(x_m)} \Wr(\mathcal P_{\mf   t_m^1}; x_m)  \bigg\}
\, d\mu^{M_1}(\mathbf x) \bigg) \epsilon_{\mf t_1^1} \wedge
\epsilon_{\mf t_2^1} \wedge \cdots \wedge \epsilon_{\mf t_{M_1}^1} \\ 
& \hspace{11cm} =
\frac{(z_1 \omega_{1})^{\wedge M_1}}{M_1!}.
\end{align*}
(The left hand side of this expression is the partition function of a
system of $M_1$ particles each of charge $q_1$ when $\beta$ is an odd
square; showing partition functions of such systems is a hyperpfaffian
was one of the goals of \cite{Sinclair:2010fk}).  

We therefore have that 
\[
Z_N(\mathbf z) = \int \bigg[ \sum_{\mathbf M}
\bigwedge_{j=1}^J
 \frac{(z_j \omega_{j})^{\wedge M_j} }{M_j!} \bigg]
 d\epsilon_{\vol} =  \int e^{\omega(\mathbf z)} d\epsilon_{\vol},
\]
as desired.  
\bibliography{bibliography}

\begin{center}
\noindent\rule{4cm}{.5pt}
\vspace{.25cm}

\noindent {\sc \small Christopher D.~Sinclair}\\
{\small Department of Mathematics, University of Oregon, Eugene OR 97403} \\
email: {\tt csinclai@uoregon.edu}
\end{center}

\end{document}